# A NOTE ON QUALIFIED MAJORITY VOTING RULES

HÉCTOR HERMIDA-RIVERA[1] • h.hermidarivera@hse.ru November 9, 2026

*This note characterizes every qualified majority voting rule (up to the designation of the reform) in environments with just two alternatives through anonymity, responsiveness, and $q$-neutrality. Crucially, the latter imposes independence of the labels of the alternatives if and only if some alternative is strictly top-ranked by at least $q$ voters. The axioms also rule out quotas not exceeding half the electorate. Thus, this note provides a characterization in the spirit of May (1952, Theorem, p. 682) that identifies the quota of each qualified majority voting rule through its degree of neutrality.*

## 1. INTRODUCTION

**Overview**. Every parliament faces the problem of choosing whether to keep or replace the status quo. If the number of legislators who strictly support either alternative (i.e., the status quo or the reform) is large, their voting rule should swap the winning alternative whenever legislators' preferences fully reverse, for their high degree of agreement would remain unchanged after the reversal. Similarly, if neither alternative gathers a large number of strict supporters, their voting rule should perhaps not swap the winning alternative whenever legislators' preferences fully reverse, for their insufficient degree of agreement would remain unchanged after the reversal. In fact, if changing a collective decision is costly, the absence of sufficiently broad support provides a reason to preserve it under preference reversal. However, sufficiently broad support for either alternative provides a reason to reverse it under preference reversal, despite those costs.

This reasoning suggests a natural neutrality-type axiom ($q$-neutrality). In environments with just two alternatives, a voting rule is neutral if and only if a complete preference reversal always triggers a swap of the winning alternative. However, under $q$-neutrality, a full preference reversal triggers a swap of the winning alternative if and only if the number of strict supporters of either alternative is sufficiently high.

The neutrality axiom was used by May (1952, Theorem, p. 682) alongside anonymity and positive responsiveness to characterize the simple majority social welfare function whenever there are just two alternatives. While legislative bodies all across the world routinely employ the simple majority voting rule, particularly important decisions often require special majorities. Given a set of $n$ voters and two alternatives, voting rules in which a given alternative is chosen if and only if at least $q > n/2$ voters strictly support it are called qualified majority voting rules. Given any such voting rule, that alternative shall be called the reform.

**Results**. Theorem 1 states that given any quota $q > n/2$, a voting rule is anonymous, responsive, and $q$-neutral if and only if it is a qualified majority voting rule with quota $q$; and that given any quota $q \leqslant n/2$, there exists no voting rule satisfying these three axioms. Further, Corollary 1 shows that anonymity can be dropped by assuming a quota $q > n/2$ from the outset.

Three features of the characterization I propose stand out. First, I provide an intuitive characterization of a very common family of voting rules that—to the best of my knowledge—has not previously been explicitly characterized as a distinct family. Second, I characterize every qualified majority voting rule up to the designation of the reform, rather than simply characterizing the entire family of such voting rules. And third, the axioms of this note naturally rule out the seldom used voting rules in which a given alternative wins if and only if it is strictly top-ranked by at least $q \leqslant n/2$ voters.[2]

[1] *Faculty of Economic Sciences, HSE University, 11 Pokrovsky Bulvar, Moscow, 109028, Russia.*

**Literature**. May's (1952) result has spurred a vast literature, some of which is closely related to this note. Fishburn (1973, Theorem 6.3, p. 67) characterizes a broader family of voting rules in which the reform wins if and only if its strict supporters reach a given positive quota. Pauly (2013, Theorem 2, p. 593) and Freixas & Samaniego (2024, Theorems 3.7, 3.8, p. 2119) characterize a broader family of voting rules in which the reform wins if and only if its strict supporters reach a given quota and outnumber the strict supporters of the status quo. Bandhu, Lahiri, & Pramanik (2020, Theorem 1) characterize the family of voting rules in which the reform wins if and only if it obtains unanimous support.

Freixas & Zwicker (2009, Theorem 4.1) and Lahiri & Pramanik (2020, Theorem 2) characterize broader families of voting rules in which the quota may depend on the number of indifferent voters. Aşan & Sanver (2006, Theorem 3.1, p. 181) and Houy (2007b, Theorem 2, p. 4) characterize slightly distinct families of voting rules in which each alternative wins if and only if its strict supporters reach a certain quota exceeding half of the full set of voters. Houy (2007a, Theorem 1, p. 20) characterizes the family of voting rules in which the reform wins if and only if its strict supporters reach a strictly positive quota of non-indifferent voters.

Llamazares (2006, Theorems 8, 13, 25, pp. 318, 320, 324) and Houy (2007c, Theorems 2–5, p. 116) characterize the family of voting rules that select the alternative with the greatest number of strict supporters if the difference between the numbers of strict supporters for each alternative is strictly greater than some quota. Finally, Jeong & Ju (2017, Theorem 3, p. 38) characterize the family of voting rules in which the reform wins if and only if its strict supporters outnumber those of the status quo and the total number of non-indifferent voters reaches a given quorum.

**Structure**. This note is organized as follows: Section 2 sets out the environment, Section 3 defines qualified majority voting rules, Section 4 introduces the axioms, and Section 5 states and proves the results.

## 2. Environment

The *environment* is a 4-tuple $(N, A, \mathcal{R}, f)$. Let $N = \{1, \dots, n\}$ be a *finite set of voters* $i, j, k, \dots$, where $n \geqslant 2$. Let $A = \{1, 2\}$ be a *set of two alternatives* $x, y$. Let

$$\mathcal{R} = \big\{(R_i)_{i \in N} \;\big|\; (\forall i)[R_i \text{ is a } \textit{weak order} \text{ on } A]\big\} \tag{1}$$

be the *set of all weak preference profiles* $R$ *on* $A$ where, given any $R$, its *strict part* $P = (P_i)_{i \in N}$ and its *indifference part* $I = (I_i)_{i \in N}$ respectively satisfy

$$(\forall i, x, y)[(xP_iy) \Leftrightarrow ((xR_iy) \wedge \neg(yR_ix))] \tag{2}$$

$$(\forall i, x, y)[(xI_iy) \Leftrightarrow ((xR_iy) \wedge (yR_ix))] \tag{3}$$

Finally, let $f : \mathcal{R} \to A$ be any *voting rule* that associates each preference profile with some alternative.

## 3. Qualified Majority Voting Rules

Given any $R$ and distinct $x, y$, let $N(x, R) = \{i \mid xP_iy\}$, $n(x, R) = |N(x, R)|$, and $N(R) = \{i \mid xI_iy\}$. Further, let $q \in \{0, 1, \dots, n\}$ be any *quota*.

[2] One notable example is the Rule of Four of the Supreme Court of the United States, under which the votes of four of its nine justices suffice to grant a *writ of certiorari*.

DEFINITION 1 — QUALIFIED MAJORITY VOTING RULE. $f$ is a *qualified majority voting rule* if and only if there exist some quota $q > n/2$ and some alternative that gets chosen exactly when it is strictly top-ranked by at least $q$ voters. Formally,

$$(\exists q > n/2)(\exists x)(\forall R)[(f(R) = x) \Leftrightarrow (n(x, R) \geqslant q)] \quad 4$$

Given any $q > n/2$ and any $x$, let $f_q^x$ be the unique qualified majority voting rule satisfying

$$(\forall R)[(f_q^x(R) = x) \Leftrightarrow (n(x, R) \geqslant q)] \quad 5$$

In $f_q^x$, alternative $x$ is the *reform*, and the other alternative is the *status quo*. Given any $q$, let the *family of qualified majority voting rules with quota* $q$ be

$$\mathcal{F}(q) = \begin{cases} \{f_q^1, f_q^2\} & \text{if } q > n/2 \\ \emptyset & \text{if } q \leqslant n/2 \end{cases} \quad 6$$

## 4. AXIOMS

Section 4 introduces three axioms: anonymity, responsiveness, and $q$-neutrality. The first two adapt the corresponding axioms of May (1952) to resolute social choice functions, while the third specifies exactly when reversing a preference profile must reverse the chosen alternative.

Let $\pi : N \to N$ be any *permutation of the voter set*. Given any $R$ and $\pi$, let $\pi R$ satisfy

$$(\forall i, x, y)[(x(\pi R)_i y) \Leftrightarrow (x R_{\pi(i)} y)] \quad 7$$

AXIOM 1 — ANONYMITY. $f$ is *anonymous* if and only if it is independent of voters' identities. Formally,

$$(\forall R, \pi)[f(R) = f(\pi R)] \quad 8$$

Let $\mathcal{F}_a = \{f \mid f \text{ is anonymous}\}$.

Given any $R$ and non-empty $S \subsetneq N$, let $R_S = (R_i)_{i \in S}$. Given any $R$ and distinct $x, y$, let $R^x$ be any profile satisfying

$$(R^x \neq R) \wedge ((\exists! i)[(R^x_{N \setminus \{i\}} = R_{N \setminus \{i\}}) \wedge (y R_i x) \wedge (x R^x_i y)]) \quad 9$$

AXIOM 2 — RESPONSIVENESS. $f$ is *responsive* if and only if the chosen alternative remains selected whenever it improves in one voter's preference while all other preferences remain fixed. Formally,

$$(\forall R, R^{f(R)})[f(R) = f(R^{f(R)})] \quad 10$$

Let $\mathcal{F}_r = \{f \mid f \text{ is responsive}\}$.

Let $\mu : A \to A$ satisfy $\mu(1) = 2$ and $\mu(2) = 1$. Given any $R$, let $\mu R$ satisfy

$$(\forall i, x, y)[(x(\mu R)_i y) \Leftrightarrow (y R_i x)] \quad 11$$

Given any $q$, let

$$\mathcal{R}(q) = \{R \mid (\exists x)[n(x, R) \geqslant q]\} \quad 12$$

AXIOM 3 — $q$-NEUTRALITY. Given any $q$, $f$ is $q$-*neutral* if and only if it is independent of the labels of the alternatives exactly when some alternative is strictly top-ranked by at least $q$ voters. Formally,

$$(\forall R)[(R \in \mathcal{R}(q)) \Leftrightarrow (\mu(f(R)) = f(\mu R))] \quad 13$$

Let $\mathcal{F}_q = \{f \mid f \text{ is } q\text{-neutral}\}$.

## 5. RESULTS

THEOREM 1. *Given any quota* $q$*, a voting rule is anonymous, responsive, and* $q$*-neutral if and only if it belongs to the family of qualified majority voting rules with quota* $q$*. Formally,*

$$(\forall q)[\mathcal{F}_a \cap \mathcal{F}_r \cap \mathcal{F}_q = \mathcal{F}(q)] \tag{14}$$

PROOF OF THEOREM 1. There are two statements to show:

1. $(\forall q)[\mathcal{F}(q) \subseteq \mathcal{F}_a \cap \mathcal{F}_r \cap \mathcal{F}_q]$,
2. $(\forall q)[\mathcal{F}_a \cap \mathcal{F}_r \cap \mathcal{F}_q \subseteq \mathcal{F}(q)]$.

STATEMENT 1. $(\forall q)[\mathcal{F}(q) \subseteq \mathcal{F}_a \cap \mathcal{F}_r \cap \mathcal{F}_q]$.

By equation 6, $\mathcal{F}(q) = \emptyset$ if $q \leqslant n/2$. Thus, consider any $q > n/2$ and $f \in \mathcal{F}(q)$. Now, the proof of Statement 1 follows from Claims 1.1 to 1.3.

CLAIM 1.1. $\mathcal{F}(q) \subseteq \mathcal{F}_a$.

The proof is direct. Consider any $R$ and $\pi$. Then, $n(1, R) = n(1, \pi R)$ and $n(2, R) = n(2, \pi R)$. By equation 5, $f(R) = f(\pi R)$. Therefore, $\mathcal{F}(q) \subseteq \mathcal{F}_a$.

CLAIM 1.2. $\mathcal{F}(q) \subseteq \mathcal{F}_r$.

The proof is direct. Consider any distinct $x, y$, any $R$ satisfying $f(R) = x$, and any $R^x$. Then, $n(x, R) \leqslant n(x, R^x)$ and $n(y, R^x) \leqslant n(y, R)$. If $f = f_q^x$, equation 5 implies that $q \leqslant n(x, R) \leqslant n(x, R^x)$; and thus, $f(R^x) = x$. If $f = f_q^y$, equation 5 implies that $n(y, R^x) \leqslant n(y, R) < q$; and thus, $f(R^x) = x$. Hence, $\mathcal{F}(q) \subseteq \mathcal{F}_r$.

CLAIM 1.3. $\mathcal{F}(q) \subseteq \mathcal{F}_q$.

The proof is direct. First, let $R \in \mathcal{R}(q)$. Since $q > n/2$, consider the unique $x$ satisfying $n(x, R) = n(\mu(x), \mu R) \geqslant q$. By equation 5, $f(\mu R) = \mu(x) = \mu(f(R))$. Second, let $R \notin \mathcal{R}(q)$. Then, $n(x, R) = n(\mu(x), \mu R) < q$ for all $x$. By equation 5, $f(\mu R) = f(R) \neq \mu(f(R))$. Hence, $\mathcal{F}(q) \subseteq \mathcal{F}_q$.

STATEMENT 2. $(\forall q)[\mathcal{F}_a \cap \mathcal{F}_r \cap \mathcal{F}_q \subseteq \mathcal{F}(q)]$.

Consider any $q$ and $f \in \mathcal{F}_a \cap \mathcal{F}_r \cap \mathcal{F}_q$. Then, the proof of Statement 2 follows from Claims 2.1 to 2.3.

CLAIM 2.1. $(q \leqslant n/2) \Rightarrow (\mathcal{F}_a \cap \mathcal{F}_r \cap \mathcal{F}_q = \emptyset)$.

The proof is by contradiction. Let $q \leqslant n/2$ and let $R$ satisfy $n(1, R) = n(2, R) = q$. Since $R \in \mathcal{R}(q)$, $q$-neutrality implies that $f(\mu R) = \mu(f(R))$. But now, since $n(x, R) = n(x, \mu R)$ for all $x$, anonymity implies that $f(R) = f(\mu R) = \mu(f(R))$, a contradiction.

CLAIM 2.2. $(q > n/2) \Rightarrow ((\forall x, R)[(n(x, R) \geqslant q) \Rightarrow (f(R) = x)])$.

The proof is by contradiction. Consider any $q > n/2$, any distinct $x, y$, and any $R$ satisfying $n(x, R) \geqslant q$ as well as $f(R) = y$. Since $R \in \mathcal{R}(q)$, $q$-neutrality implies that $f(\mu R) = \mu(f(R)) = \mu(y) = x$. Now, let $R'$ satisfy $N(y, R') = N(y, R)$ and $N(R') = N \setminus N(y, R)$. Since $f(R) = y$, successive applications of responsiveness imply that $f(R') = y$. Similarly, since $f(\mu R) = x$, successive applications of responsiveness imply that $f(\mu R') = x$. But now, since $n(x, R') = 0 < q$ and $n(y, R') = n(y, R) \leqslant n - q < q$, it follows that $R' \notin \mathcal{R}(q)$. Thus, $q$-neutrality implies that $y = f(R') = f(\mu R') = x$, a contradiction.

CLAIM 2.3. $(q > n/2) \Rightarrow ((\exists x)(\forall R)[(f(R) = x) \Rightarrow (n(x, R) \geqslant q)])$.

The proof is by contradiction. Consider any $q > n/2$. Let $\tilde{R}$ satisfy $N(\tilde{R}) = N$, let $y = f(\tilde{R})$, and let $x = \mu(y)$. Now, suppose there exists some $R$ satisfying $f(R) = x$ and $n(x, R) < q$. Let $R'$ satisfy $N(x, R') = N(x, R)$ and $N(R') = N \setminus N(x, R)$. Since $f(R) = x$, successive applications of responsiveness imply that $f(R') = x$. Since $n(x, R') = n(x, R) < q$ and $n(y, R') = 0 < q$, it follows that $R' \notin \mathcal{R}(q)$. Thus, $q$-neutrality implies that $f(\mu R') = f(R') = x$. But now, successive applications of responsiveness imply that $f(\tilde{R}) = x \neq y = f(\tilde{R})$, a contradiction. □

COROLLARY 1. *Given any quota $q > n/2$, a voting rule is responsive and $q$-neutral if and only if it is a qualified majority voting rule with quota $q$. Formally,*

$$(\forall q > n/2)[\mathcal{F}_r \cap \mathcal{F}_q = \mathcal{F}(q)] \qquad 15$$

PROOF OF COROLLARY 1. The proof follows directly from Claims 1.2 to 2.3, whose proofs do not use anonymity. □

**ACKNOWLEDGMENTS**. *I sincerely thank Róbert Somogyi, Pinaki Mandal, Toygar T. Kerman, Andrea Marietta Leina, and Christopher Stapenhurst for their helpful suggestions.* • **ARTIFICIAL INTELLIGENCE**. *I used ChatGPT 6 Astra Ultra, Google Deep Review, and Refine.ink to check the proof and improve the structure, language, and exposition of the manuscript. I reviewed the tools' outputs and take full responsibility for the submitted manuscript.* • **CONFLICT OF INTEREST**. *None.* • **DATA**. *None.* • **FUNDING**. *None.*